# Towards high-capacity quantum communications by combining wavelength- and time-division multiplexing technologies

Wen-Tan Fang,[3,+] Yin-Hai Li,[1,2,+] Zhi-Yuan Zhou,[1, 2, *] Li-Xin Xu,[3,*] Guang-Can Guo,[1,2] and Bao-Sen Shi [1, 2]

[1]*CAS Key Laboratory of Quantum Information, USTC, Hefei, Anhui 230026, China*

[2] *Synergetic Innovation Center of Quantum Information & Quantum Physics, University of Science and Technology of China, Hefei, Anhui 230026, China*

[3]*Department of Optics and Optical Engineering, University of Science and Technology of China, Hefei, Anhui 230026, China*

*[*]zyzhouphy@ustc.edu.cn; xulixin@ustc.edu.cn*

[+]*These two authors contributed equally to this article.*

**Abstract**

Optical communication systems are able to send the information from one user to another in light beams that travel through the free space or optical fibers, therefore how to send larger amounts of information in smaller periods of time is a long-term concern, one promising way is to use multiplexing of photon's different degrees of freedoms to parallel handle the large amounts of information in multiple channels independently. In this work, by combining the wavelength- and time-division multiplexing technologies, we prepare a multi-frequency-mode time-bin entangled photon pair source at different time slots by using four-wave mixing in a silicon nanowire waveguide, and distribute entangled photons into 3(time) X 14(wavelength) channels independently, which can significantly increase the bit rate compared with the single channel systems in quantum communication. Our work paves a new and promising way to achieve a high capacity quantum communication and to generate a multiple-photon non-classical state.

Quantum communication can be used to transfer quantum information between remote users through the free space or optical fibers, therefore the channel capacity is one of the most important parameters describing the communication systems. In classical optical communication systems, in order to increase the channel capacity, multiple channels are used to handle the huge amounts of information needed to be transferred independently. For example, the wavelength-division multiplexing (WDM), in which signals separated in wavelength, is used to provide many channels to transfer signals parallel; or a time-division multiplexing (TDM) is used to handle the information in different time slots independently; even recently, a spatial-division multiplexing (SDM) has been used to further improve the channel capacity [1].

Entangled photon pairs play a major role in many quantum information fields [2-5], especially in quantum communications [2]. To realize a long-distance quantum communication, a quantum repeater has to be used to overcome the problem of communication fidelity decreasing exponentially with the channel length, which requires the distribution of quantum entanglement between adjacent nodes firstly, so how to share larger amounts of entanglement in smaller periods of time is a key for achieving a high communication rate, by using multiple channels to distribute multiple entangled photon pairs independently and parallel is a very promising way. Recently, significant efforts have been devoted in this field to, for example, generate multiplexed entangled photons by WDM [6, 7] or create multiphoton entangled states [8, 9] for high-capacity and high-speed quantum information processing. Although the combination of SDM and TDM of photons, individual TDM or WDM has been used to enhance the probability of single-photon output [10-12], however, simultaneously using TDM and WDM techniques to distribute entangled photon pairs has not been realized yet.

Usually, entangled photon pairs can be generated by using spontaneous parametric down-conversion (SPDC) [13, 14] or spontaneous four-waving mixing (SFWM) [15, 16] in nonlinear crystals or waveguides. Among these materials, the silicon-on-insulator (SOI) is a very promising platform for optical quantum systems due to its high integration and CMOS compatibility. Besides, SOI-based quantum sources have much lower noise level in telecom C-band because of its much narrow Raman emission peak (105GHz) in comparison to silica (10THz) [17]. Recently, some important progresses have been achieved [16, 17] for generating different kind of entangled photon pairs.

In this work, we report in the following an experimental realization of a simultaneous time and frequency multiplexing of time-bin entangled source in a silicon nanowire waveguide (SNW). Firstly, we use temporal-multiplexed pulses to pump a SNW to generate a time-bin entangled photon pair with a broad spectrum, then we de-multiplex them in time domain by optical switches and in frequency domain by DWDM components, and distribute entangled photons into 3(time) X 14(wavelength) channels independently. By this way, we can increase the bit rate by a factor 42 compared with the single channel systems if it is applied in quantum communications. We experimentally show nearly noise-free two-photon interferences with high visibilities in 3(time) X 5 (wavelength) channels, clearly demonstrating the time-energy entanglement in each pair. Besides, we perform a four-fold Hong-Ou-Mandel (HOM) interference between photon pairs from different time slots to demonstrate the indistinguishability of photons at different time slots. Such a strategy enables the increase of dimensions of multiple entanglements modes, reveals the potential to offer a new and promising way to achieve high bit rate in quantum communication and to generate a multiple and customizable non-classical state.

**Results**
**Time- and wavelength-division multiplexing and de-multiplexing scheme.**
The principle diagram of our scheme is illustrated in Fig.1. In this scheme, a SNW is pumped by n multiplexed pulses, so entangled photon pairs with broad photon spectrum can be generated in each time slot of T1, T2 ... Tn. We use an electronic circuit to extract timing information from the pump pulse, which is subsequently used to control optical switches that actively route entangled photon pairs into *n* time slots. By merging wavelength-division multiplexing/de-multiplexing (DWDM) technique, the entangled photon pairs in each time slot are shared by *m* pairs of users. Thus, the entangled modes can be increased to *N (time) x M (wavelength)*, offering great promising for high-capacity quantum communication systems.

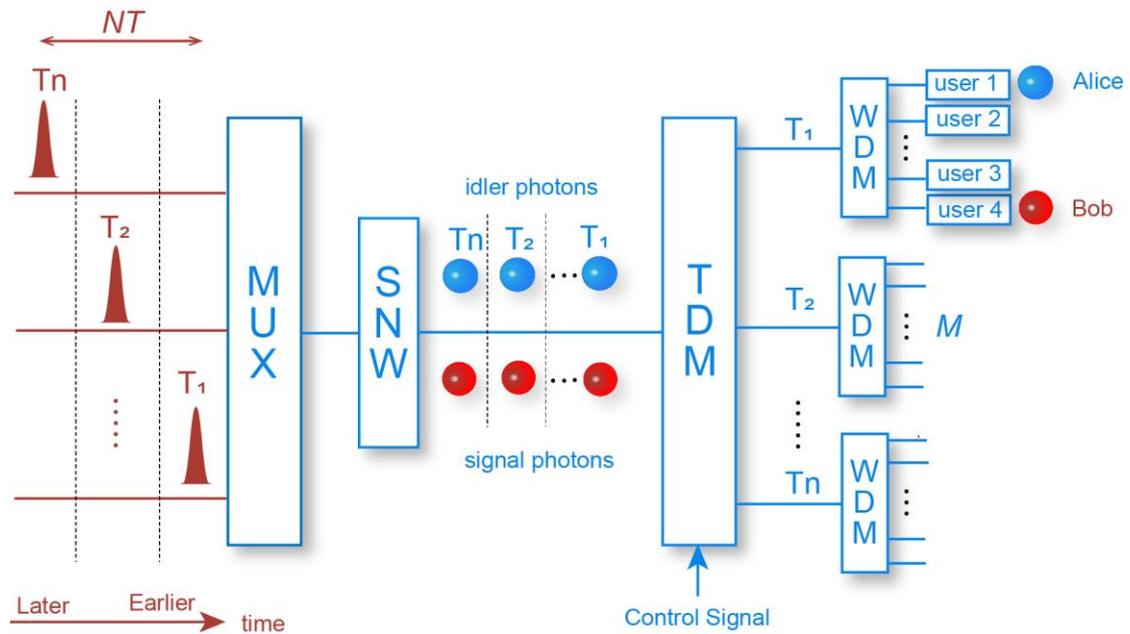

**Figure1|The principle diagram for entanglement source.** A multiplexer (MUX) assembles multiple pulses from the fundamental pulse. Multiplexed pulses pump an SNW to generate entangled photon pairs in N time slots. The entangled photon pairs are firstly split into temporal channels using a time division multiplexer (TDM). Then, the entangled photon pairs are distributed to multiusers by DWDM.

**Multiplexed time-bin entangled source**

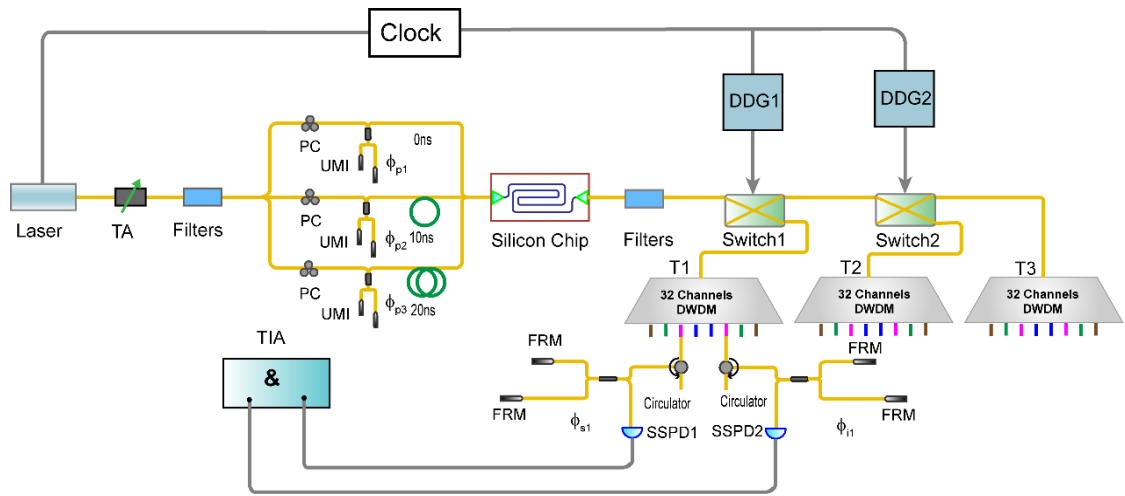

**Figure 2|Experimental setup for multiplexed time-bin entanglement source.** TA, tunable attenuator; PC, fiber polarization controller; UMI, unbalanced Michelson fiber interferometer; DDG1, DDG2, digital delay generator; FRM, Faraday rotation mirror; SSPD1 and SSPD2, superconducting single-photon detector; TIA, time interval analyzer

Fig.2 showed the entire experimental setup for producing and analyzing multiple frequency-mode time-bin entanglement. The nonlinear device employed in our experiment is a SNW with a length of 1cm and transverse dimensions of 220nm (height) x 450nm (width). The weak and anomalous dispersion of SNW enables broadband phase matching for SFWM, generating photon pairs over a broad bandwidth. Such a bandwidth would allow entanglement distribution in up to 30 standard channel pairs using high –performance multi-channel DWDM. The external pump laser is coupled to the waveguide through input-coupling gratings. And the total insertion losses from input to output port is about 10 dB, the whole loss of the system is shown in supplementary B.

The pump laser is from a mode-locked fiber laser with a repetition rate of 27.97MHz (35.75ns period) and a pulse duration 25ps. Each pulse is split into three pulses spaced by 10ns using two one-to-three fiber couplers and three optical fiber delay lines. We generated time-division multiplexed time-bin entanglements by passing three pulses through three same stabilized unbalanced Michelson fiber interferometers (UMI, 1.6ns time difference between two interfering beams) [16]. Each UMI is individually sealed in a copper box and thermally insulated from the air. The temperature of each copper box is controlled with a homemade semiconductor Peltier temperature controller with temperature fluctuations of ±2 mK.

Now we give a brief theoretical description of our time-division multiplexed time-bin entanglements. After each pulse in different channel is divided into two time bins by UMI, the pump photon is prepared in state:

$$|\psi\rangle_{p1} = \frac{1}{\sqrt{2}}(|S\rangle - e^{i\phi_{p1}}|L\rangle)$$

$$|\psi\rangle_{p2} = \frac{1}{\sqrt{2}}(|S\rangle - e^{i\phi_{p2}}|L\rangle)$$

$$|\psi\rangle_{p3} = \frac{1}{\sqrt{2}}(|S\rangle - e^{i\phi_{p3}}|L\rangle)$$

(1)

Where S and L signify the short and long arms of interferometer through which photon passes, respectively, and $\phi_{p1}, \phi_{p2}, \phi_{p3}$ is the phase difference between two interfering beams in each UMI. After the SNW, the entangled states in different time slot can be expressed as:

$$|\Phi\rangle_{T1} = \frac{1}{\sqrt{2}}(|SS\rangle - e^{i2\phi_{P1}}|LL\rangle)$$

$$|\Phi\rangle_{T2} = \frac{1}{\sqrt{2}}(|SS\rangle - e^{i2\phi_{P2}}|LL\rangle)$$

$$|\Phi\rangle_{T3} = \frac{1}{\sqrt{2}}(|SS\rangle - e^{i2\phi_{P3}}|LL\rangle)$$

(2)

A temporal de-multiplexing technique, which uses two active 1X2 optical switches and electronics synchronization control module, is used to route the multiplexed time-bin entanglements to each quantum branch channels (see fig.2), where the 1X2 optical switch is a high-speed electro-optic polarization-independent switch, made from a lithium Niobate single mode waveguide. The switching network is driven by the synchronous clock signals acquired from the mode-locked laser. We use two digital delay generators (DDG1, DDG2) to synchronize the driving signal and entangled photons.

The output from switching network is connected to three 32x100 GHz bandwidth DWDMs, belonging to the grid of the international telecommunication union (ITU). Here, photons in each pair generated in SNW are time-bin entangled. Conservation of the energy implies that photons in each pair are always produced symmetrically with respect to the center emission spectrum.

Moreover, the spectrum of the photon is broad enough to cover the entire telecom C-band, thus, the DWDM is well suited for de-multiplexing the entangled states. The corresponding channels of DWDM at equal shifts from the pump frequency are correlated. We relabel signal and idler photons S1-S14 and I1-I14 and a detail definition of the central wavelength of the ITU grid is shown in Table I in the Supplementary material. In our experiment, we fix the pump wavelength at 1550.12nm (ITU channel 31) and choose five channel pairs to experimentally characterize the quality of entanglement (see Fig. 2(a)) [16].

The signal and idler photons were each passed individually through two UMIs with an imbalance identical to that used for the pump laser. This setup allowed the measurement of the quantum interference between the signal and idler photons. To characterize the time- and wavelength-division multiplexing entanglement of the source, we selected 15 channel pairs in different time slot and different frequency and recorded quantum interference with high raw visibilities above 90% (see Fig3.a). After subtracting the background noise, the visibility was found to be above 95% in all channel pairs. If the visibility of the two-photon interference is >70.7%, the Clauster-Horne-Shimony-Holt (CHSH) inequality would be violated [26], proving the entanglement between photons. Our results clearly demonstrated the entanglement between photons in each channel pair.

To further characterize quantum states in different temporal slots, we measure the two-photon interference dependent on the relative phases of pump, signal, and in each UMI in different time channels (see Fig.2). When SFWM process (as in this work) in nonlinear media is used to generate entangled photon pairs, quantum interference is expected to be proportional to $1 - V\cos(2\phi_p - \phi_s - \phi_i)$, where $\phi_s$ and $\phi_i$ are the relative phases of the UMIs in the signal and idler ports, respectively. We obtained three groups of interference fringes (see. Fig. 3b-d) by turning the independent UIMs in three time channels, which means the phase of entanglement states in different channels can be controlled independently. As shown in Fig.3 with dashed and solid lines, the two-photon interference fringes have a period of oscillation of $\pi$ for the pump phase and $2\pi$ for the signal (idler) phase, which confirm the phase dependency in SFWM.

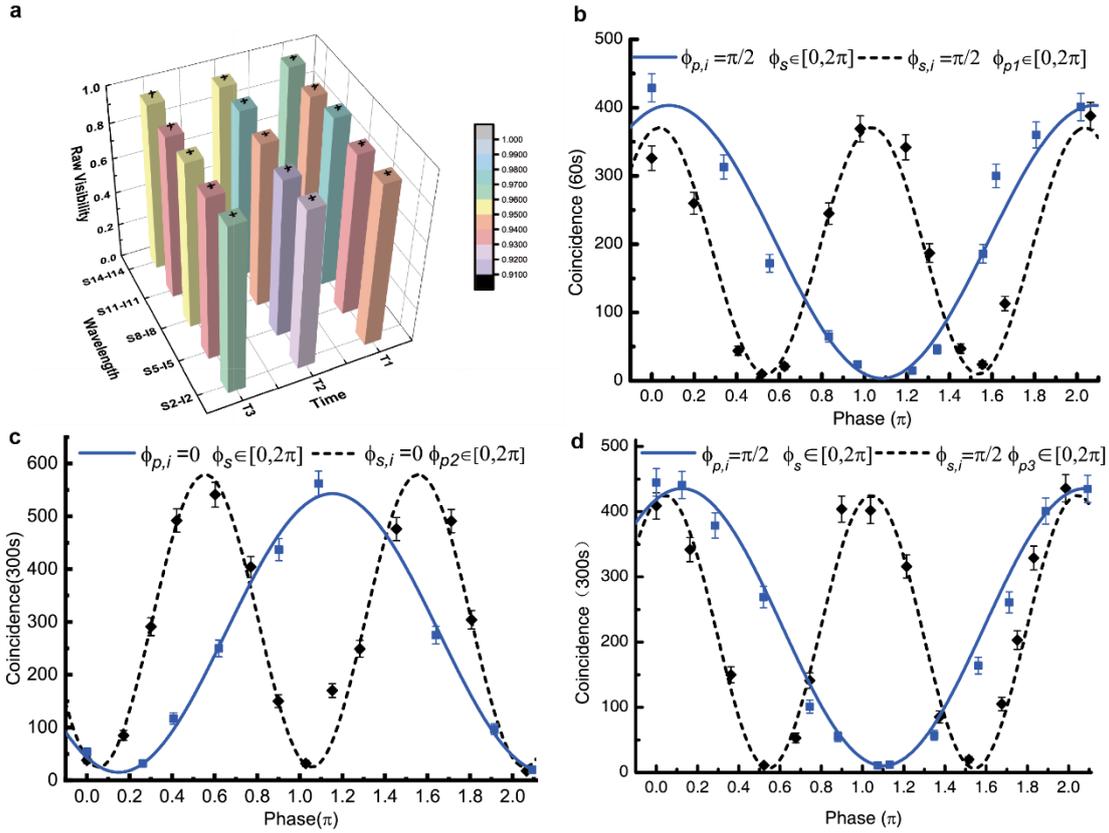

**Figure 3| Quality characterizations of time and wavelength division multiplexed entanglement source. a.** Raw visibilities of two-photon interference for 3 X 5 channel pairs. **b.** two-photon coincidence in 60s for channel S8-I8-T1 when the idler and the pump UMI phase is fixed at π/2 (solid blue line) and the idler and the signal UMI phases are fixed at π/2 (solid blue line) in channel. **c.** two photon coincidence in 300s for channel S8-I8-T2 when the idler and the pump UMI phases are fixed at 0 (solid blue line) and the idler and the signal UMI phases are fixed at 0 (solid blue line). **d.** two-photon coincidence in 300s for channel S8-I8-T3 when the idler and the pump UMI phases are fixed at π/2 (solid blue line) and the idler and the signal UMI phases are fixed at π/2(solid blue line).

**Photon pairs indistinguishability check**

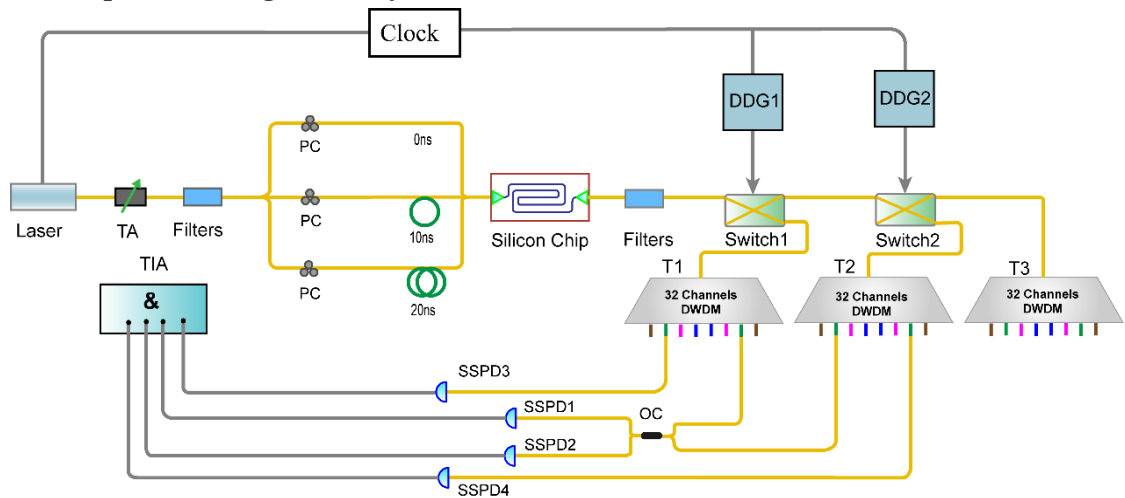

**Figure 4| Experimental setup for HOM interference between photon pairs at different time slots generated

**from the same SSW.** TA, tunable attenuator; PC, fiber polarization controller; DDG1, DDG2, digital delay generator; SSPD1 and SSPD2, superconducting single-photon detector; TIA, time interval analyzer; OC, optical coupler.

To check indistinguishability of de-multiplexed photon pairs, we performed a four-fold HOM interference experiment between photon pairs generated in different time slot (see fig.4). We removed UMIs in the first experiment and chose two groups of channel pairs for experiment (see Fig.5.). The photons with the same wavelength, to be interfered, were chosen from different time slot. An appropriate delay was applied to output photons from different channels so that they can appear in a time slot. Then idler channels were input into a 50:50 fiber coupler whose two output ports were followed by super-conducting single photon detectors (SSPD) SSPD1 and SSPD2. To ensure that the photons from the two idler channels had the same polarization, two PCs were installed before fiber coupler. The signal photons were received by SSPD3 and SSPD4 and set as trigger signals.

A higher pump powers is applied to have sufficient coincidence counts to make the statistics meaningful. With the increasing of pump power, multi-photon effect appears which reduces the visibility of the HOM dip. In the measurement, the pump power is set at a level under which coincidence to accidental coincidence ratio of 8 for both channel pairs were obtained. Fig.5 shows the obtained net four-fold coincidences as a function of optical delay. Without subtraction of any background counts, the raw visibility of the four-fold HOM dip between S14-I14-T3 and S14-I14-T2 channel pairs (S8-I8-T1 and S8-I8-T2 channel pairs) is 58.48±15.02% (55.43±3.93%), indicating that non-classical interference occurred between channel pairs in different temporal modes. The dark coincidence was measured by blocking one arm of the coupler, and summing the two results together. When subtracting the dark coincidence, we obtain net visibility of 92.9±6.2% (76.9±5.7%). These are several possible reasons for the limitation of the visibility. One is that the timing jitter caused by the relatively broad pump pulse. When the pump pulse width is comparable to or broader than the coherence time of the photon pair, we observe the timing jitter of the generated photons, which results in the temporal distinguishability [27]. Another cause of the visibility degradation is the large leakage of pump photon, which may have led to accidental coincidences caused by multiphoton emission events.

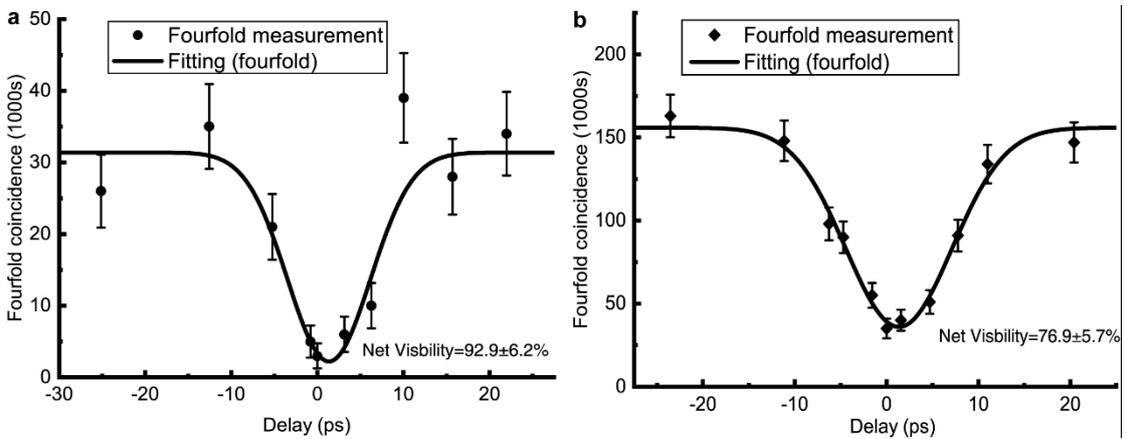

**Figure 5| a.** The measured four-fold dip between S14-I14-T3 and S14-I14-T2 channel pairs. **b.** The measured four-fold dip between S8-I8-T1 and S8-I8-T2 channel pairs

**Conclusion and outlook**

We have proposed and experimentally realized a scheme to generate a time- and wavelength-division multiplexed entangled source using a SNW, and simultaneously distributed photonic time-energy entangled photon pairs over 3 X 14 pairs of channels by combining DWDM and TDM techniques. The perfect visibilities of two-photon interference in all 3 X 5 channels clearly demonstrate the high entanglement in each channel pair. Furthermore, a four-fold HOM experiment between two photon pairs in different time slot have been reported, which demonstrates the indistinguishability between the photon pairs generated in different temporal mode. Generating photon pairs in different time slot enables an additional freedom, which makes multipartite entanglement readily available. Two-photon time-bin entangled qubits have been used successfully for linear universal quantum computation [11], and the parallel generation and processing of multiple qubits can directly enhance the information capacity. Our system's capacity can straightforwardly be scaled up by using a denser channels spacing (e.g. current commercially available 12.5GHz or 25GHz multichannel ultra-DWDMs), relied on broadband phase matching condition engineering on a single-SOI-nanowire, as it has recently been reported in [28], and by using a higher speed and lower-loss switch for increasing temporal mode. In conclusion, this work provides a road map for creating high-capacity entanglement-based quantum communication system and for generating complex quantum states, which extends significantly the ability of integrated quantum photonics.

**Acknowledgments**

This work is supported by the National Natural Science Foundation of China (NSFC) (61435011, 61525504, 61605194); the National Key Research and Development Program of China (2016YFA0302600); the China Postdoctoral Science Foundation (2016M590570); and the Fundamental Research Funds for the Central Universities.

# Supplementary Materials

**A. Definition of the wavelengths of the standard ITU grids**

The corresponding wavelengths for the correlated signal and idler photons are defined in Table S1. The bolded channels are used for characterizing the entanglement in the experiments. The pump wavelength is located at the center of channel C34.

**Table I. Definition of the wavelengths of the standard ITU grids for the signal and idler photons**

|  | DWDM channel | Wavelength ( nm ) |
|---|---|---|
| **Signal 14 - Idler 14** | **C19 - C49** | **1562.23 - 1538.19** |
| Signal 13 - Idler 13 | C20 - C48 | 1561.42 – 1538.98 |
| Signal 12- Idler 12 | C21 – C47 | 1560.61-1539.77 |
| **Signal 11 - Idler 11** | **C22 – C46** | **1559.79 – 1540.56** |
| Signal 10 - Idler 10 | C23 – C45 | 1558.98 – 1541.35 |
| Signal 9 - Idler 9 | C24 – C44 | 1558.17 – 1542.14 |
| **Signal 8 - Idler 8** | **C25 – C43** | **1557.36 – 1542.94** |
| Signal 7 - Idler 7 | C26 – C42 | 1556.56 – 1543.73 |
| Signal 6 – Idler 6 | C27 – C41 | 1555.75 – 1544.53 |
| **Signal 5 – Idler 5** | **C28 – C40** | **1554.94 – 1545.32** |
| Signal 4 - Idler 4 | C29 – C39 | 1554.13 – 1546.12 |
| Signal 3 - Idler 3 | C30 – C38 | 1553.33 – 1546.92 |
| **Signal 2 - Idler 2** | **C31 – C37** | **1552.52 – 1547.72** |
| Signal 1 - Idler 1 | C32 –C36 | 1551.72 – 1548.52 |
| **Pump** | **C34** | **1550.12** |

**B. Loss management**

We would like to give an estimation of whole detection efficiencies for photon pairs in different channels. The insertion loss of the silicon waveguide is 5.00dB for both photons, the filtering loss of the cascade DWDM filters is about 2.00dB for both photons. The insertion loss of each path of electro-optic switch is 2.5dB. The insertion loss of UMI is 4.7dB. Take into account the detection efficiency of the superconducting single-photon detector (70%, 1.5dB), the overall detection efficiency for signal photon and idler photon in T1 channel (pass through switches once) are 15.7dB and for signal photon and idler photon in T2 channel/T3 channel (pass through switches twice) are 18.2dB.

**C. More data about the photon source**

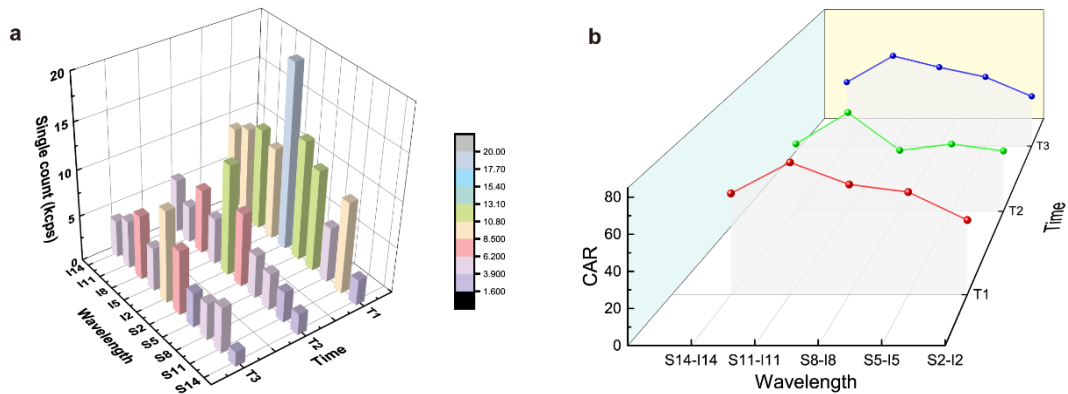

**Supplementary Figure S1 | a. single count rate for correlated channel pairs b. CARs for different correlated channel pairs from S14-I14-T1 to S2-I2-T3.**

In the single-pass configuration, we measure the single count and coincidence to accidental coincidence ratio (CAR) for 15 correlated channel pairs, when the pump power is fixed at 0.3 mw. Results are showed in Fig. S1. The difference in single count rates of the different wavelength channel arise from the different SFWM gains, Raman scattering of the signal and idler bands and nonuniform losses of different DWDM channels. The difference in single count rates between T1 and T2/T3 time channels arise from the 3dB loss of additional switch. The CAR increases when the shift in the signal and idler wavelength increase. The peak value of CAR occurs in the S11-I11 channel.